\documentclass{article}
\usepackage{times}
\usepackage{subfig}
\usepackage{textcomp}
\usepackage{graphicx}
\usepackage{booktabs}
\usepackage{amsmath}
\usepackage{nicefrac}

\newcommand{\nf}[2]{\nicefrac[\textrm]{#1}{#2}}

\title{A Bounded Confidence Approach to Understanding User Participation in Peer
Production Systems}

\author{Giovanni Luca Ciampaglia\footnote{Accepted to Socinfo 2011. The final
publication is available at www.springerlink.com}\\ciampagg@usi.ch}

\begin{document}

\maketitle

\begin{abstract}
Commons-based peer production does seem to rest upon a paradox. Although users
produce all contents, at the same time participation is commonly on a voluntary
basis, and largely incentivized by achievement of project's goals. This means
that users have to coordinate their actions and goals, in order to keep
themselves from leaving. While this situation is easily explainable for small
groups of highly committed, like-minded individuals, little is known about
large-scale, heterogeneous projects, such as Wikipedia. 

In this contribution we present a model of peer production in a large online
community. The model features a dynamic population of bounded confidence users,
and an endogenous process of user departure. Using global sensitivity analysis,
we identify the most important parameters affecting the lifespan of user
participation. We find that the model presents two distinct regimes, and that
the shift between them is governed by the bounded confidence parameter. For low
values of this parameter, users depart almost immediately. For high values,
however, the model produces a bimodal distribution of user lifespan. These
results suggest that user participation to online communities could be explained
in terms of group consensus, and provide a novel connection between models of
opinion dynamics and commons-based peer production.  
\end{abstract}

\section{Introduction}

In the past decade mass collaboration platforms have become common in several
production contexts. The term \emph{commons-based peer production} has been
coined to refer to a broad range of collaborative systems, such as those used
for producing software, sharing digital content, and organizing large knowledge
repositories, however, seem to be based upon a paradox. In wikis, there is a
link between quality and cooperation \cite{Wilkinson:2007aa}, but, at the same
time, contribution is voluntary, based on non-monetary incentives
\cite{Rafaeli:2008aa,Schroer:2009aa}.  For small teams, this might not be a
problem. In large scale wikis, where low access barriers are necessary to
attract vast masses of contributors \cite{Ciffolilli:2008aa}, and where expert
users play a crucial role in maintenance and governance
\cite{Beschastnikh:2008aa}, user retention becomes instead crucial
\cite{Goldman:2009aa}.

An established fact about participation to online groups is the
\emph{preferential behavior} of users, that is, a newcomer's long-term
participation can be predicted by the outcome of his or her early interactions
\cite{Backstrom:2007aa,Panciera:2009aa}. This could be explained in terms of
Socialization theory \cite{Choi:2010aa}, as users assess the willingness of the
community to accept them and vice versa.  It is also true, however, that quality
assessment of the produced contents, and in particular comparison of the
objectives of an individual with those of the community, is important in
determining user participation \cite{Lin:2006aa}.  This could be explained as a
form of day-to-day coordination or group consensus taking place among editors
\cite{Kittur:2010aa}.  

In this paper we study user participation as a collective social phenomenon
\cite{Castellano:2009aa}. Other models of peer-production have been proposed
already, for example for social information filtering platforms
\cite{Hogg:2009aa}. Here, we draw specifically from the modeling work on models
of social influence under bounded confidence
\cite{Deffuant:2001aa,Hegselmann:2002aa}. 

Let us consider a community of users engaged in editing a collection of pages,
e.g. Wikipedia. Pages are denoted by a certain number of features upon which
users can find themselves in agreement or not. For example, let us consider the
writing style of pages. Users try to modify pages according to their objectives,
i.e.  using their own style. At the same time, by interacting with contents,
users can be also influenced by the style of other users. This reciprocal
influence, however, happens only to a certain extent, that is, only when user
and page (that is, their styles) are similar enough. Vandals, to
illustrate with the same example, might not be interested in learning the
encyclopedic writing style. In the context of social psychology this phenomenon
is known as \emph{bounded confidence}, and is regarded as a general feature of
human communication within groups that try to reach consensus
\cite{Hegselmann:2002aa}. It can be also seen as a form of herding in that
people are influenced by the social context they are in \cite{Raafat:2009aa}.

The population of users in our model is dynamic, with user departure determined
endogenously by the social influence process. Although others have already
studied Deffuant's model to a dynamic population \cite{Carletti:2008aa}, here we
explicitly link the process of social influence to user participation.

We implemented these ideas in an agent-based model of a peer production system.
In this model, several factors affect the behavior of agents, such as user
activity, content popularity, and community growth. To understand what factors
are truly important for the resulting dynamics of user participation, we
performed a factor screening using global sensitivity analysis.

\subsection{Related work}

The subject of user participation in mass collaboration systems has been already
touched by several authors, for example on social networking sites
\cite{Leskovec:2008aa}, and knowledge sharing platforms \cite{Yang:2010aa}. A
``momentum'' law has been proposed for the distribution of user life edits of
inactive users \cite{Wilkinson:2008aa}. The distribution of user account
lifespans has been shown to decay with a heavy tail, and a power-law model has
been proposed after this observation \cite{Grabowski:2010aa}. Empirical data
from Wikipedia, however, seem to support a super-position of different regimes
\cite{Ciampaglia:2010aa}; a feature of the model we present here is indeed a
bimodal distribution of user lifespans. In the context of wikis and other free
open source initiatives some authors have used survival analysis to outline the
diffences between different communities, \cite{Ortega:2009ab} but this modeling
technique is not suited to understand the connection between social influence,
group coordination, and user retention. We advocate the need to explicitly model
such processes explicitly.

The paper is organized as follows: in Sec. \ref{sec:model} we introduce our
model of peer production; in Sec. \ref{sec:methods} we briefly describe global
sensitivity analysis and Gaussian Processes, the two statistical techniques we
used for the factor screening study; in Sec. \ref{sec:results} we present our
main results and we discuss them in Sec. \ref{sec:discussion}.

\section{An agent-based model of commons-based peer production}
\label{sec:model}

In this section we introduce our model of peer production. While we make
explicit use of the terminology of wiki platforms (e.g. ``users'' who ``edit
pages'') we stress that ours is a general model of consensus building in a
dynamic bipartite population, and not merely a description of a wiki platform.
We also stress that in our model the state of agents may not necessarily
represent an opinion in the classic sense of other studies of opinion dynamics,
i.e. extremes of the spectrum do not necessarily denote -- say -- political
extremism, nor we speak of ``moderates'' to identify the center of the opinion
space.

To keep things simple, we consider only the unidimensional case, i.e. the state
of an agent is a scalar number in the interval $\left[ 0,1 \right]$. We denote
with $x\left( t \right)$ the state of a generic user at time $t$ and with
$y\left( t \right)$ the state of a generic page.

The interaction rule between a user and a page captures the dynamics of social
influence. Let us imagine that at time $t$ a user edits a page. Let
$\mu\in\left[ 0,\nf{1}{2} \right]$ be the speed (or uncertainty) parameter and
$\varepsilon\in\left[ 0,1 \right]$ the confidence \cite{Lorenz:2007ab}. If
$|x\left( t \right) - y\left( t \right)| < \varepsilon$ then: 

\begin{eqnarray} \label{eq:mod:1} x\left( t \right) &\gets& x\left( t \right) +
    \mu\left( y\left( t \right) - x\left( t \right)\right) \\ \label{eq:mod:2}
    y\left( t \right) &\gets& y\left( t \right) + \mu\left( x\left( t \right) -
    y\left( t \right)\right) \end{eqnarray}

\noindent else, if $|x\left( t \right) - y\left( t \right)| \ge \varepsilon$, we
allow only Eq. \eqref{eq:mod:2} to take place with probability $r$. This
addition to the bounded confidence averaging rule reflects the fact that, in
peer production systems, users often deal with content they do not agree with
without being influence by it, as when a vandalized page is reverted to a
previous, non-vandalized revision (also known as rollback).

Different pages can reflect different topics and hence receive attention from
users based on their popularity. We employ a simple reinforcement mechanism to
model this. Let $c_p \ge 0$ be a constant. If $m_t$ is the number of edits a
page has received up to time $t$, then the probability of it being selected at
that time will proportional to $m_t + c_p$. When $c_p\rightarrow \infty$, pages
will be chosen for editing with uniform distribution, regardless of the number
of edits they have received. Hence, we can study the impact of of content
popularity in user participation by setting $c_p$ to a small or large value. Of
course users do not always choose to edit an existing page. Sometime, a user can
decide to create a new page. We model this by considering a rate of new page
creations $\rho_p$. Whenever a new page is created, its state $y$ is equal to
the state $x$ of creator. Creators are chosen at random among existing users.

In order to model user participation, the population of users is dynamic. First,
we consider an input rate of new users $\rho_u$, whose state is chosen at random
within the interval $\left[ 0,1 \right]$. Second, we consider a inhomogeneous
departure rate that depends on the experience of users. Let us consider a
generic user at time $t$ and let us denote with $n_t$ the number of edits he (or
she) did up to $t$, and with $s_t$ the number of these edits that resulted in
the application of Eq. \eqref{eq:mod:1}. Let $c_s \ge 0$ be a constant and
$r(t)$ be the ratio

\begin{equation} r\left( t \right) = \frac{s_t + c_s}{n_t + c_s}
    \label{eq:mod:4} \end{equation}

\noindent The rate of departure $\lambda_d\left( t \right)$ is then defined as:

\begin{equation} \lambda_d\left( t \right) = \frac{r\left( t \right)}{\tau_0} +
    \frac{1 - r\left( t \right)}{\tau_1} \label{eq:mod:3}
\end{equation}

\noindent with $\tau_0 \gg \tau_1$ time scale parameters. Depending on the value
of $r\left( t \right)$, the expected lifetime $\left< \tau \right>$ will
interpolate between two values: $\left< \tau \right> = \tau_0$ (long lifetime)
for $r\left( t \right) = 1$, $\left< \tau \right> = \tau_1$ if $r\left( t
\right)= 0$ (short lifetime). If $c_s \rightarrow \infty$, we recover a
homogeneous process with rate $\tau_0^{-1}$, so we can set $c_s$ to control how
sensitive the departure rate is to unsuccessful interactions.

\section{Evaluation Methods}
\label{sec:methods}

\subsection{Computer Code Emulation via Gaussian Processes}

Although we can perform the statistical evaluation of our peer production model
using directly the computer simulator, this approach is not desirable, as
evaluation of the computer code can be quite time consuming. We rely instead on
emulation of the computer code output. We use a Gaussian Process (GP) as a
surrogate model of the average lifetime $\left<\tau\right>$ of users in our peer
production system. Gaussian processes (or Gaussian Random Functions, GRF) are a
supervised learning technique used for functional approximation of smooth
surfaces and for prediction purposes: see \cite{Santner:2003aa} for the
application of GP to computer code evaluation.

Given input sites $\boldsymbol\Theta_\textrm{obs} = \left( \boldsymbol\theta_1,
\boldsymbol\theta_2, \dotsc, \boldsymbol\theta_N \right)$ we can evaluate our
model as specified above, and obtain observations of the average user lifetime
$T_\textrm{obs} = \left( \tau_1, \tau_2, \dotsc, \tau_N \right)$. Based on these
observations, we wish to predict the value of $\tau$ at an untested input site
$\theta$, i.e.  $\tau\left( \theta \right)$. With a GP, this value is
$\hat\tau\left( \theta \right)= \textrm E\left[ \tau\left( \theta \right) \,|\,
\Theta_\textrm{obs} \right]$; the uncertainty in the prediction, that is,
$\textrm{Var}\left[ \hat\tau\left( \theta \right) \right]$, is equal to
$\textrm{Var}\left[ \tau\left( \theta \right) \,|\, \Theta_\textrm{obs}
\right]$. With it we can compute a confidence interval that characterizes the
uncertainty of the prediction of $\tau$ based on training data $\left(
\Theta_\textrm{obs}, T_\textrm{obs} \right)$.

There are several strategies for selecting the input sites $\Theta_\textrm{obs}$
at which we will run our computer simulator. Here we choose to employ a uniform,
space-filling design generated via Latin Hypercube Sampling (LHS) because it
yields better error bounds than those produced with uniform random sampling
\cite{McKay:1992aa}. The space-filling requirement is attained using a
\emph{maximin} design. A maximin design is any collection of points $\Theta$
that maximizes the minimum distance between points:
\begin{math}\label{eq:gsa:expdes:lhs:3}
    \max_{\boldsymbol\Theta}{\min_{i<i'} \big\|\boldsymbol\theta_i - \boldsymbol \theta_{i'} \big\|}
\end{math}

\subsection{Global Sensitivity Analysis}

A computational or mathematical model is comprised usually of a number of
parameters, or factors, which are meant to affect in some way its output, or
response. Hence, in general, a model can be thought as a mapping between factors
(input) and responses (output). One might be interested in the problem of
quantifying how much output ``variability'' in this mapping can be apportioned
to each of the inputs. Global Sensitivity Analysis (GSA) is a set of statistical
techniques used to get an answer to this problem. See \cite{Saltelli:2004aa} for
a primer on GSA.

One application of GSA is \emph{factor screening}. The ranking of parameters is
usually done by computing the sensitivity indices of each input parameter
(factor). There are various techniques for computing the sensitivity indices,
each with its own properties and assumptions. In this study we computed
sensitivity indices by decomposing the output variance of our surrogate model.
We used other techniques as well, namely partial correlation coefficients and
standardized regression coefficients, and they gave concordant results. We
choose to report here only the results of the decomposition of variance because
it applies more naturally to non-linear models like ours.

The method we use was proposed by Sobol\textasciiacute{} and is based on the
analysis of variance (\textsc{anova}) \cite{Sobol:2001aa}. The idea is to
decompose the variance of the output in several components that are attributable
to independent factors, in our case the parameters of the model.

Let us assume that the space of parameters is $\big[0,1\big]^d$, where $d$ is
the number of parameters. Sobol\textasciiacute{} proposes to write the output
$Y$ as:

\begin{multline}\label{eq:gsa:glob:var:1}
    Y\big(\theta_1,\dotsc,\theta_d\big) = 
    Y_0 + \sum_{i=1}^d Y_i\big(\theta_i\big) + 
    \sum_{1\le i < j\le d} Y_{i,j}\big(\theta_i, \theta_j\big) + 
    \\ + 
    \dotsb + 
    Y_{1,2,\dotsc,d} \big( \theta_1, \theta_2, \dotsc, \theta_d \big)
\end{multline}

\noindent and shows that this decomposition is unique under the assumption that
components are orthogonal and have zero mean. In Eq. \eqref{eq:gsa:glob:var:1},
$Y_0 = E\left[Y\right]$, $Y_i\big(\theta_i\big)$ is the \emph{main} effect of
parameter $\theta_i$, $Y_{i,j}(\theta_i, \theta_j)$ is the 2-way
\emph{interaction} effect between the $i$-th and $j$-th parameters ($i \neq j$),
and so on. Each summand is computable from suitable integrals. For example the
main effect of $Y_i$ is:

\begin{equation}\label{eq:gsa:glob:var:4}
    Y_i(\theta_i) = \int_0^1 \dotsi \int_0^1 Y(\theta_1, \dotsc, \theta_d)
    d\boldsymbol\theta_{\neg i} - Y_0
\end{equation}

\noindent where with $\boldsymbol\theta_{\neg i}$ when mean the reduced
parameter vector obtained by considering all parameters except $\theta_i$.
Similar formulas can be obtained for higher order effects. Let us now consider
the variances of the summands of Eq. \eqref{eq:gsa:glob:var:1}. We can decompose
$\sigma^2$, the total variance of $Y$, as:

\begin{equation}\label{eq:gsa:glob:var:5}
    \sigma^2 = \sum_{i=1}^d \sigma^2_i + \sum_{1\le i < j \le d}\sigma^2_{i,j}
    + \dotsb + \sigma^2_{1,2,\dotsc,d}
\end{equation}

The sensitivity indices proposed by Sobol\textasciiacute{} are obtained by
standardizing all summands of Eq. \eqref{eq:gsa:glob:var:5}, obtaining:

\begin{equation}\label{eq:gsa:glob:var:6} 
    1 = \sum_{i=1}^d M_i + \sum_{1\le i < j \le d} C_{i,j} + \dotsb +
    C_{1,2,\dotsc d} 
\end{equation}

\noindent $M_i$ is the \emph{main sensitivity index} of parameter $\theta_i$,
$C_{i,j}$ is the \emph{two-way interaction index} between $\theta_i$ and
$\theta_j$, etc. Two quantities are of interest for assessing the importance of
a parameter: the already cited main sensitivity index $M_i$; and the \emph{total
interaction index} $T_i$, which is defined as the sum of all terms that involve
parameter $\theta_i$:

\begin{equation}\label{eq:gsa:glob:var:7}
    T_i = \sum_{j\neq i}C_{i,j} + \sum_{\substack{1\le j < k\le d \\ j,k \neq
    i}} C_{i,j,k} + \dotsb + C_{1,2,\dotsc d}
\end{equation}

\section{Results}
\label{sec:results}

\subsection{Simulation scenario}

\begin{table}[tH]
    \centering
    \caption{Parameters settings for global sensitivity analysis.}
    \begin{tabular*}{\textwidth}{@{\extracolsep{\fill}}llcclr@{\extracolsep{0pt}}}
        \toprule
        Parameter & Variable name & Symbol & Value(s) & Unit & Distribution \\
        \midrule
        Const. popularity    & const\_pop     & $c_p$         & $(0, 100)$        &               & uniform \\
        Const. successes     & const\_succ    & $c_s$         & $(0, 100)$        &               & uniform \\
        Confidence           & confidence     & $\varepsilon$ & $(0, \nf{1}{2})$  &               & uniform \\
        Daily edit rate      & daily\_users   & $\lambda_e$   & $(1, 20)$         & day           & uniform \\
        Daily rate of pages  & daily\_pages   & $\rho_p$      & $(1, 20)$         & $\nf{1}{day}$ & uniform \\
        Daily rate of users  & daily\_edits   & $\rho_u$      & $(1, 20)$         & $\nf{1}{day}$ & uniform \\
        Initial no. of pages &                & $N_p$         &                   &               & see Subsec. \ref{sec:gsa:res:tra} \\
        Initial no. of users &                & $N_u$         &                   &               & see Subsec. \ref{sec:gsa:res:tra} \\
        Long lifetime        & long\_life     & $\tau_0$      & $(10, 100)$       & day           & uniform \\
        Rollback probability & rollback\_prob & $r$           & $(0, 1)$          &               & uniform \\
        Short lifetime       & short\_life    & $\tau_1$      & $(\nf{1}{24}, 1)$ & day           & uniform \\
        Simulation time      &                & $T$           & $1$               & year          &         \\
        Speed                & speed          & $\mu$         & $(0, \nf{1}{2})$  &               & uniform \\
        Transient time       &                & $T_0$         & $2$               & year          &         \\ 
        \bottomrule
    \end{tabular*}
    \label{tab:gsa:res:sce:1}
\end{table}

Table \ref{tab:gsa:res:sce:1} lists all parameters of the model, together with
simulation settings. Two quantities have been held fixed: simulation time, and
transient time. Two other parameters, the initial number of users $N_u$ and the
initial number of pages $N_p$, are determined after a transient, see Subsec.
\ref{sec:gsa:res:tra} below. All remaining parameters, instead, were assigned a
range of values. To sum up, we had an input space of 10 independent dimensions.

We chose the long ($\tau_0$) and short ($\tau_1$) user lifetimes to range in
non-overlapping intervals corresponding to different time scales, consistently
with empirical observations of user participation from Wikipedia
\cite{Ciampaglia:2010aa}. The value of the simulation time $T$ was chosen so
that a simulation would comprise more than one generation of long-term users.

Intervals for event rates such as the daily rate of edits ($\lambda_e$), of new
user arrivals ($\rho_u$), and of new page creations ($\rho_p$), were chosen
looking at plausible values from the public statistics on the Wikipedia
project.\footnote{These statistics are freely available on
\href{http://stats.wikimedia.org}{http://stats.wikimedia.org}.}. These
parameters have a strong influence on simulation time, therefore ranges for them
were set trying to strike a balance between exhaustiveness of the sensitivity
analysis and simulation wall clock time.  

The choice of ranges for the constant popularity term ($c_p$) and for the
constant successes term ($c_s$) was a bit more problematic. To our knowledge,
none of them has ever been studied before in the context of peer production
communities. We settled for ranges we deemed would be large enough for our
purposes. 

Finally, the opinion dynamics parameters. It is clear that $\mu < \nf{1}{2}$.
Regarding the confidence $\varepsilon$, the literature on bounded confidence
models in one dimension suggests that for $\varepsilon > \nf{1}{2}$ the dynamics
of consensus does not change noticeably. This should apply also to the dynamics
of user participation in our model. We ran some simulations of the average
lifetime, and found confirmation to this intuition. We thus restricted
$\varepsilon$ to the interval $(0, \nf{1}{2})$.

\subsection{Transient}\label{sec:gsa:res:tra}

\begin{figure}[t]
    \centering
    \subfloat[][]{
        \includegraphics[width=0.5\columnwidth]{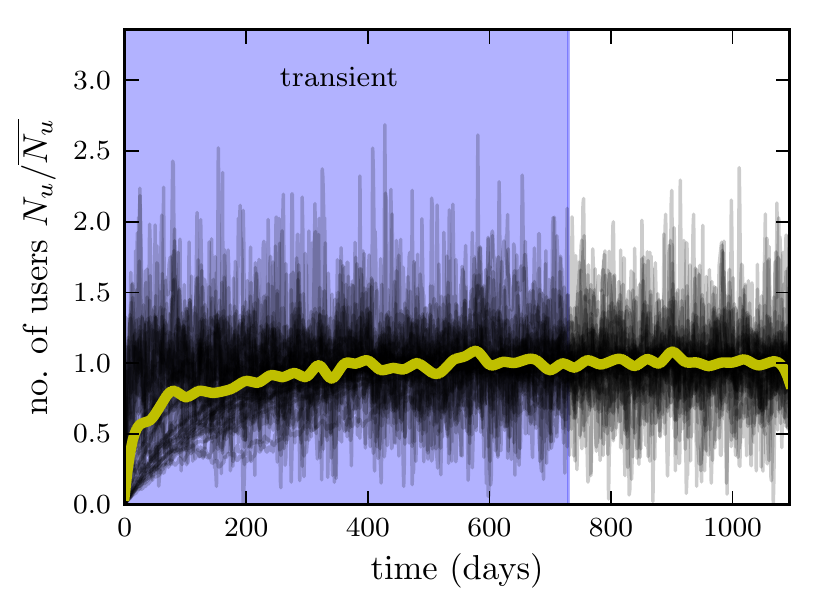}
        \label{fig:res:tra:1a}
    }
    \subfloat[][]{
        \includegraphics[width=0.5\columnwidth]{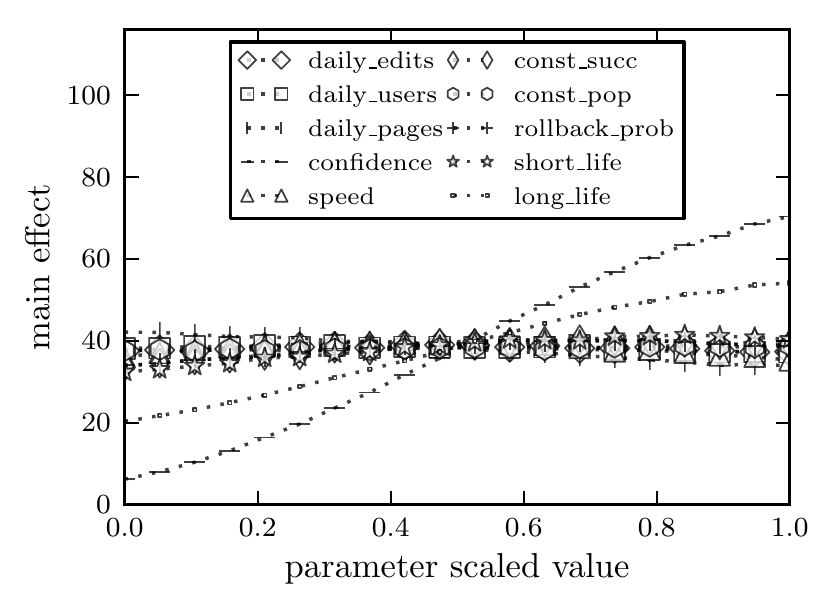}
        \label{fig:res:gsa:1b}
    }
    \caption{\subref{fig:res:tra:1a}: transient time determination.
    \subref{fig:res:gsa:1b}: main effects plot.}\label{fig:res:1}
\end{figure}

Transient duration $T_0$ was determined empirically: we plotted the daily number
of users $N_u\left( d;\, \boldsymbol\theta \right)$, $d=1,2,\dots$, for various
values of the parameters $\boldsymbol\theta$ and chose $T_0$ as the time after
which all curves look stationary. Figure \ref{fig:res:tra:1a} reports the
results of this exercise. In the figure, the shaded region corresponds to the
transient interval $\left(0, T_0\right)$. The value of $T_0$ is 730 days. The
values of $\boldsymbol\theta$ were taken from a maximin LHD with 50 points. Each
curve is scaled by its average value $\overline{N_u}\left( \boldsymbol\theta
\right)$ computed over the interval $d\in\left[731, 1095\right]$. The yellow
solid line is a B-spline fit of 50 evenly spaced observations of the expected
scaled number of users $\nicefrac{N_u}{\overline N_u}$, and serves as a guide
for the eye.

During the transient phase we did not record any data, so that the estimation of
$\tau$, on which the sensitivity our analysis is based, did not reflect the
dynamics of opinion formation during the transient. 

\subsection{Factor screening via global sensitivity analysis}

We sampled a maximin Latin Hypercube Design (LHD) with 50 points using the
intervals listed in Tab. \ref{tab:gsa:res:sce:1}. To sample a decent maximin
design, we generated $10^4$ hypercubes at random and selected the one that
maximized Eq. \eqref{eq:gsa:expdes:lhs:3}. We computed the average user lifetime
$\left<\tau\left( \boldsymbol\theta \right)\right>$ by running 10 replications
for any $\boldsymbol\theta$ and averaging the values obtained.

\begin{figure}[t] 
    \centering
    \includegraphics{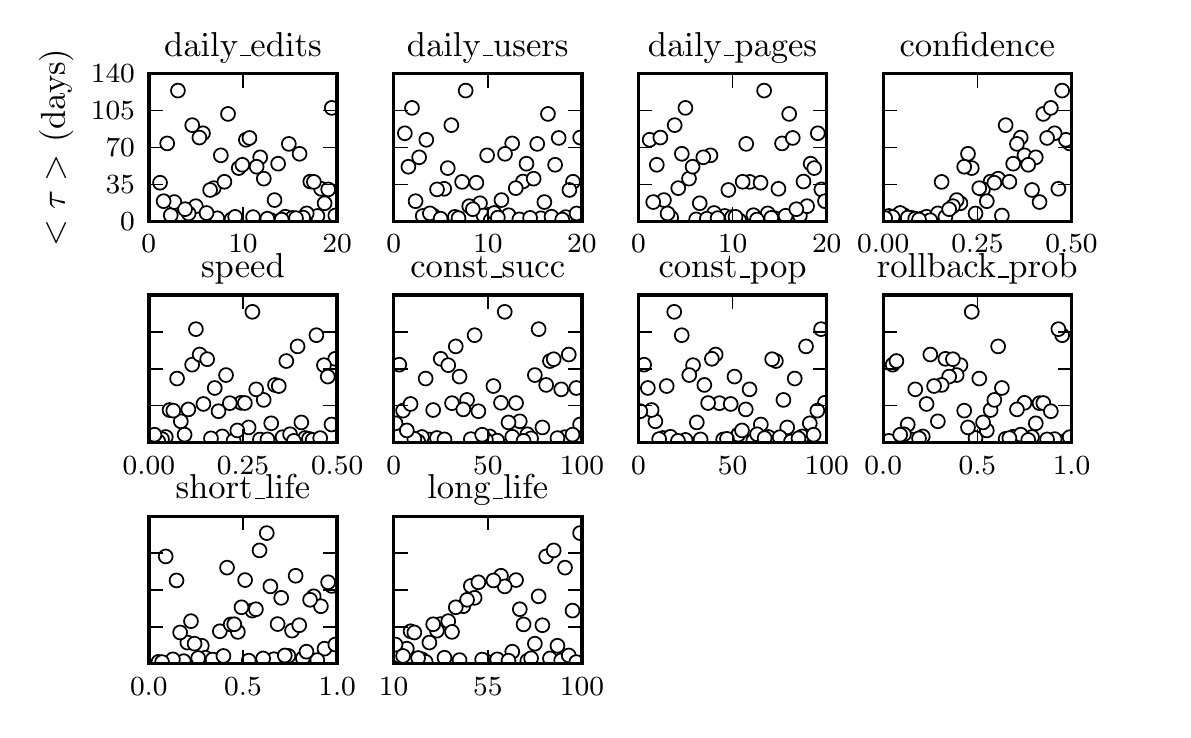} 
    \caption{Scatter plots of $\big<\tau\big>$ versus $ \boldsymbol\theta =
    \left(\lambda_e, \rho_u, \rho_p, \varepsilon, \mu, c_s, c_p, r, \tau_0,
    \tau_1 \right)$. Error bars (standard error of the mean lifetime computed
    over 10 realization) are all smaller than the data
    points.}
    \label{fig:gsa:res:fs:1}
\end{figure}

We first plotted the values of the response variable $\tau$ versus each input
parameter to check visually for any linear trend. Scatter plots are shown in
Fig. \ref{fig:gsa:res:fs:1}. A multiple linear regression gave a coefficient of
determination $R^2 = 0.83$. However, no clear trend emerges from the plots for
all parameters except for the confidence $\varepsilon$ and the long lifetime
$\tau_0$. For the latter, something similar to a linear trend can be seen,
whereas for the other the relationship looks more of sigmoidal type. We tried
fitting a sigmoid function to $\tau$ as a function of $\varepsilon$. The result
of a K-S test (p-value $<3.5\times10^{-4}$) rejected the normality of the
residuals, and therefore led us to exclude a sigmoid model as a possible
functional form of $\tau\left( \varepsilon \right)$.

Next, we fitted a GP emulator to the average user lifetime data, using the open
source machine learning toolkit from the SciKits collection\footnote{Home page:
\href{http://scikit-learn.sourceforge.net/}{http://scikit-learn.sourceforge.net/}.}
We then discarded the simulator and used $\hat\tau\left( \boldsymbol\theta
\right)$ in lieu of it. To compute the sensitivity indices we used the Winding
Stairs (WS) method, a resampling technique proposed in \cite{Jansen:1994aa}. We
computed main ($M_i$) and total interaction ($T_i$) effect indices for each
parameter ($i=1\dots10$) using a WS matrix with $10^4$ rows. The results are
shown in Tab. \ref{tab:gsa:res:fs:3}. 

\begin{table}[t]
    \caption{Variance decomposition. Winding Stairs sample size $10^{4}$ rows, total variance $635.365$ days\textsuperscript{2}.}
    \centering
    \begin{tabular*}{\textwidth}{@{\extracolsep{\fill}} c r@{.}@{\extracolsep{0pt}}l@{\extracolsep{\fill}} r@{.}@{\extracolsep{0pt}}l}
        \toprule
        Parameter & \multicolumn{2}{c}{$M_i$} & \multicolumn{2}{c}{$T_i$} \\
        \midrule
        $\lambda_e$   & -0&002 & 0&014\\
        $\rho_u$      & -0&003 & 0&02\\
        $\rho_p$      & 0&003  & 0&027\\
        $\varepsilon$ & 0&65   & 0&73\\
        $\mu$         & -0&004 & 0&03\\
        $c_s$         & 0&004  & 0&03\\
        $c_p$         & -0&005 & 0&016\\
        $r$           & -0&005 & 0&026\\
        $\tau_1$      & 0&002  & 0&03\\
        $\tau_0$      & 0&18   & 0&23\\
        \bottomrule
    \end{tabular*}
    \label{tab:gsa:res:fs:3}
\end{table}

The total variance $\hat\sigma^2$ was also computed from $\boldsymbol W$ (each
column of a WS matrix is an independent sample). The WS method yields better
estimates of the total interaction effects than other methods 
\cite{Chan:2000aa}, so we impute the presence of some slightly negative values
of $M_i$ to the uncertainty in the estimation of the total output variance
$\sigma^2$ and to the presence of factors with almost null total effect.

Only two factors have a $T_i > 3 \%$. These are the confidence $\varepsilon$,
and the long term lifetime $\tau_0$. We explored further the individual
contribution of each parameter in the output variance by looking at the main
effect plots. These are plots of $Y\left( \theta_i \right)$ as a function of
$\theta_i$, and can be obtained evaluating Eq. \eqref{eq:gsa:glob:var:4} using
Monte Carlo averaging and the GP emulator. To facilitate comparison of the
different parameter ranges, in Fig. \ref{fig:res:gsa:1b} we plotted the main
effect as a function of the scaled parameter value. Figure \ref{fig:res:gsa:1b}
shows that $\rho_p$ and $\tau_1$ have a slight effect on user lifetime too, the
first negative and the second positive.

The difference between $T_i$ and $M_i$ is the fraction of variance that is only
due to interactions between $\theta_i$ and any other parameter or groups of
parameters. For $\varepsilon$ this difference is $0.08$ and for $\tau_0$ it is
$0.05$. Summed up together, this residual interaction effect amounts to almost
three quarters (77\%) of the total interaction effects from all remaining
parameters. Thus we expect $\varepsilon$ and $\tau_0$ to have some interesting
interactions with other parameters. We explored two-way interactions
systematically using two-way interaction plots, which are the 3D counterparts of
the curves of Fig. \ref{fig:res:gsa:1b}. 

\begin{figure}[t]
    \begin{center}
        \includegraphics{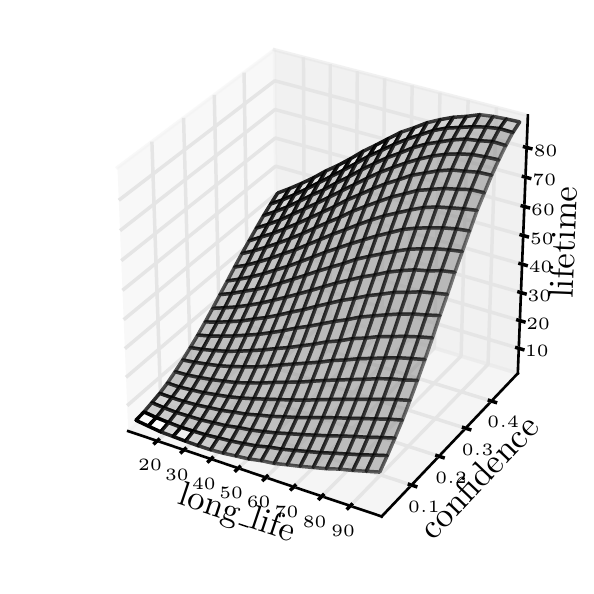}
        \includegraphics{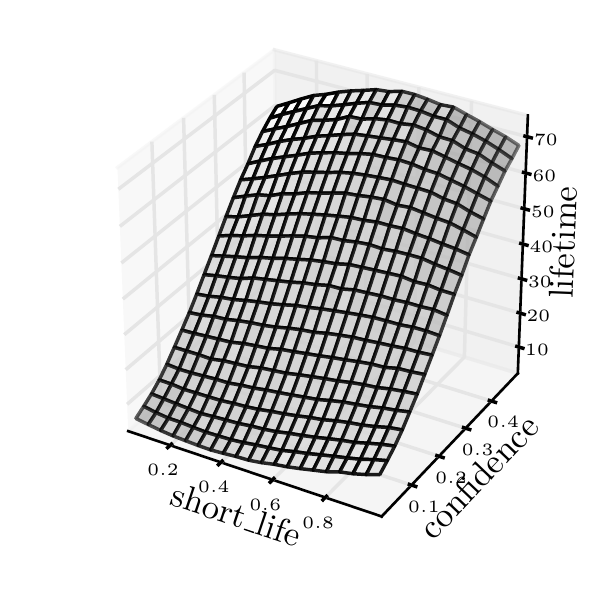}
        \includegraphics{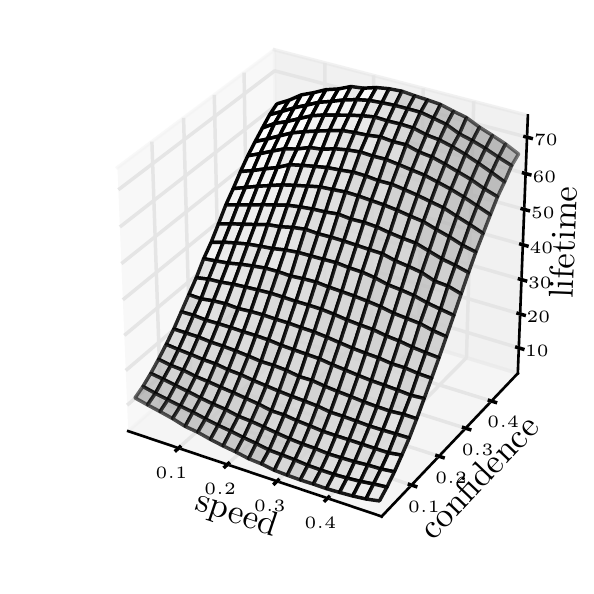}
        \includegraphics{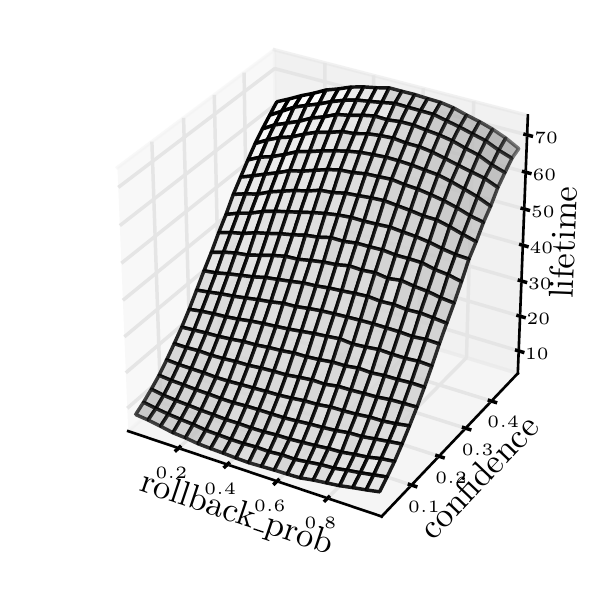}
    \end{center}
    \caption{Two-way interaction plots}
    \label{fig:gsa:res:fs:3}
\end{figure}

\begin{figure}[t]
    \begin{center}
        \includegraphics{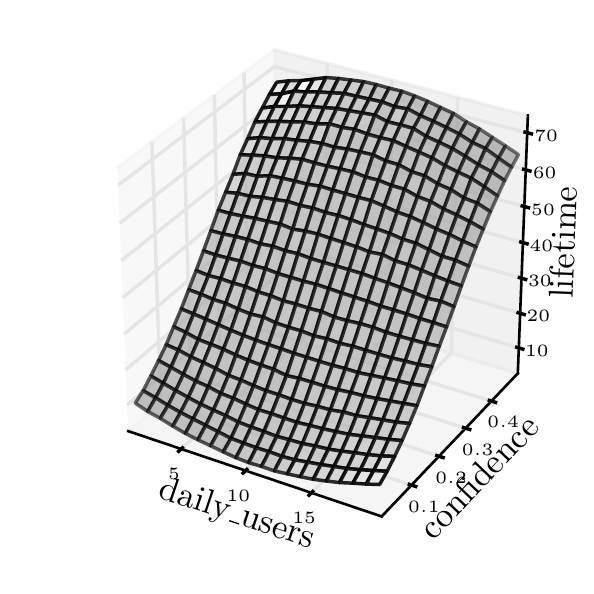}
        \includegraphics{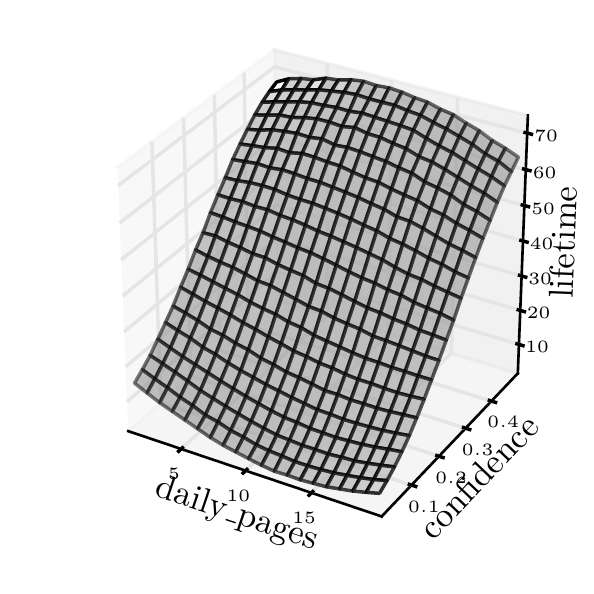}
        \includegraphics{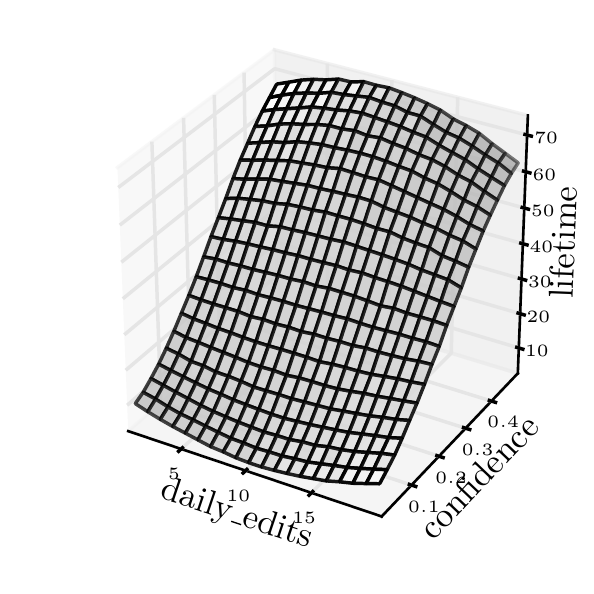}
    \end{center}
    \caption{Two-way interaction plots (cont'd)}
    \label{fig:gsa:res:fs:4}
\end{figure}

Given two parameters $\theta_i$ and $\theta_j$, with $i\ne j$, we computed
$Y_{i,j}\big(\theta_i,\theta_j)$: we evaluated Eq. \eqref{eq:gsa:glob:var:4} in
a similar way, this time holding fixed the values of two parameters instead of
one. Here we report the results on the interaction between $\varepsilon$ and
other parameters, included $\tau_0$. The plots are shown in Fig.
\ref{fig:gsa:res:fs:3} and \ref{fig:gsa:res:fs:4}. 

Almost all parameters show just a weak interaction with $\varepsilon$, which
occurs at low ($\varepsilon<0.1$) and high ($\varepsilon>0.4$) values of it.
Only the pair $\{\varepsilon,\tau_0\}$ shows a significant degree of
interaction.

\subsection{User lifetime distribution}

\begin{figure}[t]
    \begin{center}
        \includegraphics[width=0.45\columnwidth]{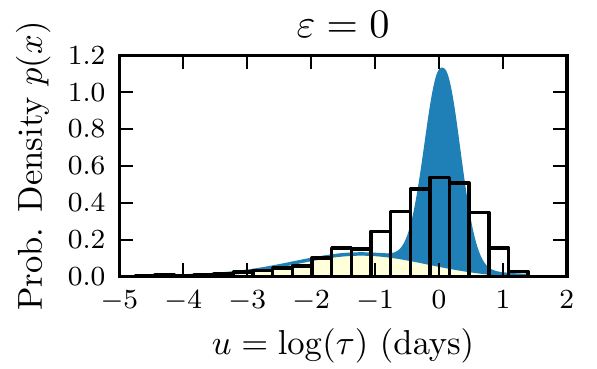}
        \includegraphics[width=0.45\columnwidth]{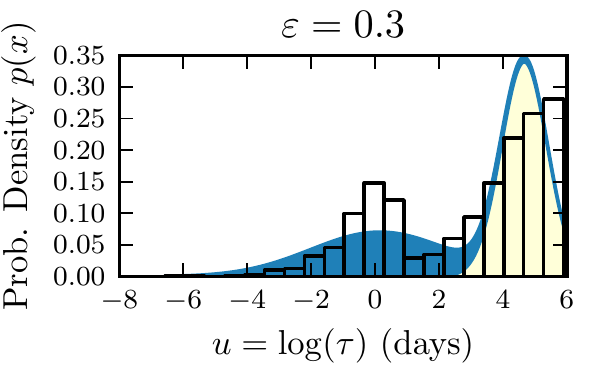}
    \end{center}
    \caption{GMM fit of log-lifetime of user accounts in two different runs of
    the model. For $\varepsilon > \varepsilon_c$ a bi-modal pattern is a clear
    feature of user participation.}
    \label{fig:res:lt:1}
\end{figure}

Previous studies on continuous opinion dynamics under bounded confidence show
that, as $\varepsilon$ grows, the population of agents undergoes a gradual
change from a regime with no consensus, to a regime of total consensus with a
single cluster \cite{Deffuant:2001aa,Hegselmann:2002aa}. In our model this shift
must reflect somehow in the average user lifetime, but what shape the user
lifetime distribution takes during it? The findings from the previous section
let us restrict the field of study to just two parameters of the original ten,
namely $\varepsilon$ and $\tau_0$. In this section we focus only on them, and
try to understand what is the actual distribution of user lifetimes, by
simulating from our model.

We performed simulations holding fixed the user lifetime parameters ($\tau_0 =
100$ days and $\tau_1 = 1$ hour), while changing the value of $\varepsilon$. The
values of all other parameters were fixed to the midpoints of the respective
ranges listed in Tab. \ref{tab:gsa:res:sce:1}. We computed the log-lifetime $u =
\log\left( \tau \right)$ and fitted a 2-components Gaussian Mixture Model (GMM)
to $u$. Figure \ref{fig:res:lt:1} reports the result of the fitting, showing the
densities of the individual components using stacked area plots. We report here
only two values of $\varepsilon$, $\varepsilon = 0$ and $\varepsilon = 0.3$,
which is a value greater than the threshold for consensus in Deffuant's model,
to show the difference between the two regimes. 

\section{Discussion}
\label{sec:discussion}

In this section we discuss the main findings of the present study. We presented
an agent-based model of user participation in a peer-production community. We
model participation as a bounded confidence consensus process, where users
modify content according to their objectives and skills (represented by a
continuous state $x$), and are in turn indirectly influenced by the rest of the
community. We use global sensitivity analysis to study the importance of the
model's parameters in explaining the average user lifetime. The first
interesting -- and rather surprising -- finding is that, as shown in Tab.
\ref{tab:gsa:res:fs:3}, of the overall ten parameters of the model, only two
affect the average user lifetime in a considerable way. This is interesting
because it suggests that several other factors like content popularity, user
community growth, and user activity rate, are not as important as the general
level of ``tolerance'' of the community (given by the confidence $\varepsilon$)
in affecting the process of group consensus. Moreover, interaction plots show
that relevant interactions occur between $\varepsilon$ and $\tau_0$: this
confirms the intuition that the role of $\tau_0$ is to set the support of the
distribution of $\tau$, and that $\varepsilon$ acts as a switch, controlling the
transition from a regime where only short-term forms of participation are
possible, due to the low rate of successful user-page interactions, to a
consensus regime where a cluster of long-term users is able to emerge.

Of course, the results from the factor screening should be also viewed in light
of our simulation setup. We decided to focus on a stable community, where the
number of users $N_u$ is stationary, and not on the initial phase of community
formation. Plausibly, during this transient phase other parameters, such as the
speed $\mu$, and the rollback probability $r$, might have more importance in
determining the span of user participation.

The second interesting finding is about the actual distribution of user
participation, which is markedly bimodal. From Fig. \ref{fig:res:lt:1} it is
possible to appreciate, for $\varepsilon = 0.3$, a clear subdivision in two
groups of users based on their participation span. We can see also a subdivision
for $\varepsilon = 0$, which is probably related to the fact that $c_s = 50$ in
that setup. Although we did not perform a proper model calibration, this finding
is encouraging, as previous studies on the distribution of user accounts
lifetime in Wikipedia have shown a similar bimodal pattern
\cite{Ciampaglia:2010aa}. 

In general, both findings show that agent-based model can be studied through the
systematic use of simulations and computer code emulation, and provide a novel
connection between model of opinion dynamics, whose study has been so far
notoriously lacking on the empirical side
\cite{Castellano:2009aa,Sobkowicz:2009aa}, and peer production.

\subsubsection{Acknowledgments.}

Alberto Vancheri and Paolo Giordano for the insightful discussions; the
anonymous reviewers, for the suggestions on how to improve the manuscript; the
conference organization, for their generous financial support.

\bibliographystyle{splncs03}
\bibliography{socinfo2011.bib}

\end{document}